\documentclass[12pt,english]{article}
\usepackage[latin1]{inputenc}
\usepackage{amsmath}
\usepackage{amssymb}

\makeatletter
\usepackage{babel}
\makeatother
\begin{document}
\begin{center}{\huge Canonical Formulation }\par\end{center}{\huge \par}

\begin{center}{\huge of }\par\end{center}{\huge \par}

\begin{center}{\huge pp-waves }\par\end{center}{\huge \par}

\vspace{1cm}

\begin{center}{\large Herbert BALASIN\renewcommand{\thefootnote}{\fnsymbol{footnote}}\footnote[1]{email: hbalasin@tph.tuwien.ac.at}\renewcommand{\thefootnote}{\arabic{footnote}}}\par\end{center}{\large \par}

\begin{center}\textit{\small Institut für Theoretische Physik, TU-Wien,
}\\
\textit{\small Wiedner Hauptstraße 8-10, 1040 Wien, }\\
\textit{\small AUSTRIA}\par\end{center}{\small \par}

\begin{center}{\large and }\par\end{center}{\large \par}

\begin{center}{\large Peter C. AICHELBURG\renewcommand{\thefootnote}{\fnsymbol{footnote}}\footnote[7]{email: aichelp8@univie.ac.at}\renewcommand{\thefootnote}{\arabic{footnote}}}\par\end{center}{\large \par}

\begin{center}\textit{\small Institut für Theoretische Physik, Universität
Wien, }\\
\textit{\small Boltzmanngasse 5, 1090 Wien, }\\
\textit{\small AUSTRIA}\par\end{center}{\small \par}

\vspace{1cm}

\begin{abstract}
We construct a Hamiltonian formulation for the class of plane-fronted
gravitational waves with parallel rays (pp-waves). Because of the
existence of a light-like Killing vector, the dynamics is effectively
reduced to a 2+1 evolution with {}``time'' chosen to be light-like.
In spite of the vanishing action this allows us to geometrically identify
a symplectic form as well as dynamical Hamiltonian, thus casting the
system into canonical form.
\end{abstract}
\vspace{1cm}

\newpage{}

\section*{Introduction}

Plane-fronted gravitational waves with parallel rays (pp-waves)are
considered as highly idealized wave phenomena which have been extensively
studied in General Relativity and related areas. These geometries
are characterized by the existence of a covariantly constant null
vector field $p^{a}$. In a by now classical work Jordan, Ehlers and
Kundt \cite{JEK} have given a complete classification of the pp-vacuum
solutions to the Einstein field equations in terms of their spacetime
symmetries.

The special class of the so called impulsive pp-waves (which were
excluded in \cite{JEK} but treated in \cite{AiBa2,AiBa3}), i.e.
geometries which are flat space everywhere except of a single null-hypersurface
generated by $p^{a}$ have been found to describe both the ultra-relativistic
(null-limit) of (stationary) black holes as well as the gravitational
field generated by massless particles \cite{AS}. This has led to
a semiclassical investigation of particle scattering at ultrahigh
(Planckian) energies within these backgrounds which displays amplitudes
similar to those appearing in String theory \cite{tHooft,Verl}. Also,
pp-waves belong to the class of algebraic special solutions of Petrov
type N. Moreover, all invariants formed from the curvature tensor
vanish identically. This property has made them a candidate as an
exact background for the consistent propagation of strings \cite{Pol}.
Due to the above mentioned richness it seems worthwhile to investigate
a possible quantization of this family of geometries. From the canonical
point of view due to the frozen degrees of freedom such a quantization
should yield a midi-superspace model \cite{Ku}. Unfortunately the
vanishing of the action for the whole class of pp-waves does not allow
a {}``straightforward'' Hamiltonian formulation, which relies on
Legendre transformation of the Lagrangian. However, upon a careful
analysis of the equations of motion, we succeed in the construction
of a symplectic form as well as a Hamiltonian (vector field) which
generates the evolution. The investigation of this structure will
be the aim of the present work.

Our work is organized as follows: After stating our conventions, we
briefly review the timelike situation in terms of a Gaussian decomposition
of an arbitrary metric. This section is mainly devoted to familiarize
the reader with the concepts used in the lightlike situation. Section
two derives a similar evolution formulation for the lightlike (pp-wave)
setting, which yields an effectively 2+1 dimensional situation. In
section three we discuss the propagation of the constraints of the
pp-wave system. Finally, section four casts the dynamical system into
Hamiltonian form, as a pre-requisite for quantization.

\section*{0 ~Conventions}

Our conventions with respect to metric and covariant derivatives follow
closely that of Wald \cite{Wald} (with $a,b\dots$referring to abstract
indices)\begin{gather}
\nabla_{a}g_{bc}=0\,\,\mbox{and}\,\,[\nabla_{a},\nabla_{b}]f=0\nonumber \\
{}[\nabla_{a},\nabla_{b}]v^{b}=R^{b}\,_{mab}v^{m}\nonumber \\
R_{ab}=R^{c}\,_{acb},\,\, R=g^{ab}R_{ab}\end{gather}
The signature of the (spacetime) metric is taken to be $(-+++).$
In terms of (normalized) tetrads we have\begin{equation}
g_{ab}=\eta_{\alpha\beta}e_{a}^{\alpha}e_{b}^{\beta}\qquad g^{ab}=\eta^{\alpha\beta}E_{\alpha}^{a}E_{\beta}^{b}\end{equation}
where $E_{\alpha}^{a}$ and $e_{a}^{\alpha}$ denote dual frames.
The Cartan structure relations for the spin-connection $\omega^{\alpha}\,_{\beta a}$,
the Riemann two-form $R^{\alpha}\,_{\beta ab}$ and the Ricci one-form
$R^{\alpha}\,_{a}$ become

\begin{gather}
de^{\alpha}=-\omega^{\alpha}\,_{\beta}e^{\beta}\nonumber \\
R^{\alpha}\,_{\beta}=d\omega^{\alpha}\,_{\beta}+\omega^{\alpha}\,_{\gamma}\omega^{\gamma}\,_{\beta}\nonumber \\
R_{\alpha}=E_{\beta}\lrcorner R^{\beta}\,_{\alpha}\nonumber \\
R=E_{\beta}\lrcorner R^{\beta}\end{gather}
where the skew (wedge) product in the above relations is implicitly
understood. The hook $\lrcorner$ denotes the contraction of a $p$-form
with a vector field

\section{Einstein equations in Gaussian coordinates}

In order to gain some familiarity with the approach used for pp-waves
let us begin with the well-known $3+1$decomposition of the Einstein
equations in terms of Gaussian coordinates

\begin{equation}
ds^{2}=-dt^{2}+h_{ij}(x,t)dx^{i}dx^{j}\end{equation}
where the $t=const$ surfaces denote the spacelike slices of the Gaussian
coordinate system. Using a canonically adapted tetrad \begin{equation}
e^{\alpha}=(dt,\tilde{e}^{i}(x,t))\qquad\qquad E_{\alpha}=(\partial_{t},\tilde{E}_{i}(t,x))\end{equation}
the corresponding connection is derived from the structure equations
(in the expression above we have explicitly exhibited the parametric
$t$- dependence, whose derivatives will be denoted by a dot in the
following, e.g. $\partial_{t}\tilde{e}^{i}=\dot{\tilde{e}}^{i}$)
\begin{eqnarray}
 &  & d\tilde{e}^{i}=-\tilde{\omega}^{i}\,_{j}\tilde{e}^{j}+dt\dot{\tilde{e}}^{i}=-(\tilde{\omega}^{i}\,_{j}+F^{i}\,_{j}dt)\tilde{e}^{j}-K^{i}\,_{j}\tilde{e}^{j}dt\nonumber \\
 &  & \omega^{i}\,_{j}=\tilde{\omega}^{i}\,_{j}+F^{i}\,_{j}dt,\quad\omega^{i}\,_{t}=K^{i}\,_{j}\tilde{e}^{j},\end{eqnarray}
where we have decomposed $\dot{\tilde{e}}^{t}$ with respect to $\tilde{e}^{i}$
and split the corresponding coefficient matrix $\tilde{E}_{i}\lrcorner\dot{\tilde{e}}^{i}$
into its symmetric and antisymmetric part respectively\[
K^{i}\,_{j}=\frac{1}{2}(\tilde{E}_{j}\lrcorner\dot{\tilde{e}}^{i}+\tilde{E}^{i}\lrcorner\dot{\tilde{e}}_{j})\quad F^{i}\,_{j}=\frac{1}{2}(\tilde{E}_{j}\lrcorner\dot{\tilde{e}}^{i}-\tilde{E}^{i}\lrcorner\dot{\tilde{e}}_{j}).\]
From this we derive the components of the Riemann 2-form\begin{eqnarray}
R^{i}\,_{j} & = & d\omega^{i}\,_{j}+\omega^{i}\,_{l}\omega^{l}\,_{j}+\omega^{i}\,_{t}\omega^{t}\,_{j}\nonumber \\
 & = & \tilde{R}^{i}\,_{j}+dt\dot{\tilde{\omega}}^{i}\,_{j}+\tilde{D}F^{i}\,_{j}dt+K^{i}\,_{l}K_{jm}\tilde{e}^{l}\tilde{e}^{m},\nonumber \\
R^{i}\,_{t} & = & d\omega^{i}\,_{t}+\omega^{i}\,_{j}\omega^{j}\,_{t}\nonumber \\
 & = & \tilde{D}K^{i}\,_{j}\tilde{e}^{j}+(\dot{K}^{i}\,_{j}+K^{i}\,_{l}K^{l}\,_{j}-K^{i}\,_{l}F^{l}\,_{j}+F^{i}\,_{l}K^{l}\,_{j})dt\tilde{e}^{j},\end{eqnarray}
and finally the Ricci 1-forms\begin{eqnarray}
R_{t} & = & E_{i}\lrcorner R^{i}\,_{t}\nonumber \\
 & = & (\tilde{D}_{i}K^{i}\,_{j}-\tilde{D}_{j}K)\tilde{e}^{j}-(\dot{K}+K^{i}\,_{j}K^{j}\,_{i})dt,\nonumber \\
R_{i} & = & E_{t}\lrcorner R^{t}\,_{i}+E_{j}\lrcorner R^{j}\,_{i}\\
 & = & (K_{ij}+KK_{ij}-K_{il}F^{l}\,_{j}+F_{il}K^{l}\,_{j})\tilde{e}^{j}+\tilde{R}_{i}-dt\tilde{E}_{j}\lrcorner\dot{\tilde{\omega}}^{j}\,_{i}+\tilde{D}_{j}F^{j}\,_{i}dt.\nonumber \end{eqnarray}
where \[
\tilde{D}v^{i}=\tilde{d}v^{i}+\tilde{\omega}^{i}\,_{j}\tilde{v}^{j}\]
denotes the exterior covariant derivative relative to the $t=const.$
surfaces. The Einstein equations reduce to\begin{eqnarray}
\tilde{D}_{i}K^{i}\,_{j}-\tilde{D}_{j}K & = & 0,\nonumber \\
\dot{K}+K^{i}\,_{j}K^{j}\,_{i} & = & 0,\nonumber \\
\dot{K}{}_{ij}+KK_{ij}-K{}_{il}F^{l}\,_{j}+F_{il}K^{l}\,_{j}+\tilde{R}_{ij} & = & 0,\end{eqnarray}
where we took into account that \[
\tilde{D}_{i}K^{i}\,_{j}-\tilde{D}_{j}K-\tilde{D}_{i}F^{i}\,_{j}=\tilde{E}_{i}\lrcorner\dot{\tilde{\omega}}^{i}\,_{j}\]
which follows from taking the {}``time'' derivative of three-dimensional
structure equation $\tilde{d}\tilde{e}^{i}=-\tilde{\omega}^{i}\,_{j}\tilde{e}^{j}$.
The first equation and the difference between the second and the trace
of the third equation are just the well-known momentum and Hamilton
constraints of General Relativity respectively\begin{eqnarray}
\tilde{D}_{i}K^{i}\,_{j}-\tilde{D}_{j}K & = & 0,\nonumber \\
K^{i}\,_{j}K^{j}\,_{i}-K^{2}-\tilde{R} & = & 0.\end{eqnarray}
which are constraints on the initial data whereas dynamics is contained
in\begin{equation}
\dot{K}{}_{ij}+KK_{ij}-K{}_{il}F^{l}\,_{j}+F_{il}K^{l}\,_{j}+\tilde{R}_{ij}=0.\end{equation}
The arbitrariness of the choice of triad $\tilde{e}_{a}^{i}$ reflects
itself in the appearance of the antisymmetric {}``field-strength''
$F^{i}\,_{j}$. However, if we re-express the equations in terms of
the 3-metric $h_{ab}=\delta_{ij}e^{i}\,_{a}e^{j}\,_{b}$ we find\begin{eqnarray}
\dot{h}_{ab} & = & \delta_{ij}(\dot{\tilde{e}}^{i}\,_{a}\tilde{e}^{j}\,_{b}+\tilde{e}^{i}\,_{a}\dot{\tilde{e}}^{j}\,_{b})\nonumber \\
 & = & \delta_{ij}((K^{i}\,_{l}-F^{i}\,_{l})\tilde{e}^{l}\,_{a}\tilde{e}^{j}\,_{b}+\tilde{e}^{i}\,_{a}(K^{j}\,_{l}-F^{j}\,_{l})\tilde{e}^{l}\,_{b})\nonumber \\
 & = & 2K_{ij}\tilde{e}^{i}\,_{a}\tilde{e}^{j}\,_{b}=2K_{ab}.\end{eqnarray}
Together with the other equation we therefore have a decomposition
of the Einstein equations into dynamical and constraint equations.\begin{eqnarray}
\dot{h}_{ab}=2K_{ab} &  & D_{a}K^{a}\,_{b}-D_{b}K=0\nonumber \\
\dot{K}_{ab}=-KK_{ab}-\tilde{R}_{ab} &  & K^{a}\,_{b}K^{b}\,_{a}-K^{2}-\tilde{R}=0\end{eqnarray}
(where $D_{a}$ denotes the Levi-Civita derivative of $h_{ab}$)

\section{2+1(+1) decomposition of pp-waves}

In this section we describe an analogous (quasi-Gaussian) decompositon
for pp-wave geometries, i.e. metrics characterized by the existence
of a covariantly constant null vector-field\begin{equation}
ds^{2}=-2dudv+\sigma_{ij}(x,u)dx^{i}dx^{j}\end{equation}
Since $p^{a}=\partial_{v}^{a}$ generates a Killing symmetry we are
dealing effectively with a 2+1 decomposition of a system dimensionally
reduced. However, the {}``time'' direction is chosen to be lightlike.
This null direction is geometrically singled out by being orthogonal
to the (arbitrarily) chosen (spacelike) 2-slices. The canonically
adapted tetrad is given by\begin{equation}
e^{\alpha}=(du,dv,\tilde{e}^{i}(u,x))\qquad\qquad E_{\alpha}=(\partial_{u},\partial_{v},\tilde{E}_{i}(x,u))\end{equation}
(As in the timelike-case, a dot will denote the derivative with respect
to the {}``time''-parameter $u$) \begin{eqnarray}
 &  & d\tilde{e}^{i}=-\tilde{\omega}^{i}\,_{j}\tilde{e}^{j}+du\dot{\tilde{e}}^{i}=-(\tilde{\omega}^{i}\,_{j}+F^{i}\,_{j}du)\tilde{e}^{j}-K^{i}\,_{j}\tilde{e}^{j}du\nonumber \\
 &  & \omega^{i}\,_{j}=\tilde{\omega}^{i}\,_{j}+F^{i}\,_{j}du\quad\omega^{i}\,_{u}=K^{i}\,_{j}\tilde{e}^{j}\end{eqnarray}
where we used the decomposition of $\tilde{E}_{i}\lrcorner\dot{\tilde{e}}^{j}:=\tilde{E}_{i}^{a}\dot{\tilde{e}}_{a}^{j}$
into symmetric and antisymmetric parts, respectively\[
K^{i}\,_{j}=\frac{1}{2}(\tilde{E}_{j}\lrcorner\dot{\tilde{e}}^{i}+\tilde{E}^{i}\lrcorner\dot{\tilde{e}}\,_{j})\qquad F^{i}\,_{j}=\frac{1}{2}(\tilde{E}_{j}\lrcorner\dot{\tilde{e}}^{i}-\tilde{E}^{i}\lrcorner\dot{\tilde{e}}\,_{j}).\]
 Therefore the non-vanishing components of the curvature 2-form are\begin{eqnarray}
R^{i}\,_{j} & = & d\omega^{i}\,_{j}+\omega^{i}\,_{l}\omega^{l}\,_{j}\nonumber \\
 & = & \tilde{R}^{i}\,_{j}+du\dot{\tilde{\omega}}^{i}\,_{j}+\tilde{D}F^{i}\,_{j}du\nonumber \\
R^{i}\,_{u} & = & d\omega^{i}\,_{u}+\omega^{i}\,_{j}\omega^{j}\,_{u}\\
 & = & \tilde{D}K^{i}\,_{j}\tilde{e}^{j}+(\dot{K}^{i}\,_{j}+K^{i}\,_{l}K^{l}\,_{j}-K^{i}\,_{l}F^{l}\,_{j}+F^{i}\,_{l}K^{l}\,_{j})du\tilde{e}^{j}\nonumber \\
where &  & \tilde{D}\tilde{v}^{i}=\tilde{d}\tilde{v}^{i}+\tilde{\omega}^{i}\,_{j}\tilde{v}^{j}\nonumber \end{eqnarray}
Together with identity\[
(\tilde{D}K^{i}\,_{j}-\tilde{D}F^{i}\,_{j})\tilde{e}^{j}+\dot{\tilde{\omega}}^{i}\,_{j}\tilde{e}^{j}=0\]
one easily obtains the Ricci one-form\begin{eqnarray}
R_{u} & = & E_{i}\lrcorner R^{i}\,_{u}\nonumber \\
 & = & (\tilde{D}_{i}K^{i}\,_{j}-\tilde{D}_{j}K)\tilde{e}^{j}+(\dot{K}+K^{i}\,_{j}K^{j}\,_{i})du\nonumber \\
R_{i} & = & E_{v}\lrcorner R^{v}\,_{i}+E_{j}\lrcorner R^{j}\,_{i}\nonumber \\
 & = & \tilde{R}_{i}+(-\tilde{E}_{j}\lrcorner\dot{\tilde{\omega}}^{j}\,_{j}+\tilde{D}_{j}F^{j}\,_{i})du\nonumber \\
 & = & \tilde{R}_{i}+(\tilde{D}_{j}K^{j}\,_{i}-\tilde{D}_{i}K)du\end{eqnarray}
in accordance with the symmetry of the Ricci tensor. Imposing the
vacuum equations results in\begin{eqnarray}
\tilde{D}_{i}K^{i}\,_{j}-\tilde{D}_{j}K & = & 0\nonumber \\
\tilde{R}_{i} & = & 0\nonumber \\
\dot{K}+K^{i}\,_{j}K^{j}\,_{i} & = & 0\end{eqnarray}
Switching back to the metric representation we find\begin{equation}
\dot{\sigma}_{ab}=(\delta_{ij}\tilde{e}^{i}\,_{a}\tilde{e}^{j}\,_{b})^{\bullet}=2K_{ab}\end{equation}
which once again gives a split into evolution and constraint equations\begin{eqnarray}
\dot{\sigma}_{ab}=2K_{ab} &  & D_{a}K^{a}\,_{b}-D_{b}K=0\nonumber \\
\dot{K}+K^{a}\,_{b}K^{b}\,_{a}=0 &  & \tilde{R}_{ab}=\frac{1}{2}\sigma_{ab}\tilde{R}=0\end{eqnarray}
(where $D_{a}$ denotes the Levi-Civita connection associated with
$\sigma_{ab}$)\\
The Ricci constraint entails the flatness of the two-dimensional sections
which in turn allows the explicit solution of the {}``momentum''
constraint, via Fourier-transforms \begin{equation}
K^{a}\,_{b}=D^{a}D_{b}\frac{1}{D^{2}}K,\end{equation}
where the action of the inverse of $D^{2}$ is given by the corresponding
convolution with the Green-function of the two-dimensional Laplace
operator.

\section{Propagation of the constraints}

In order to show that {}``time'' evolution respects the constraint
equations we will consider first the variation of the Ricci-scalar\begin{equation}
\delta\tilde{R}=-\delta\sigma^{ab}\tilde{R}_{ab}+D_{a}D_{b}\delta\sigma^{ab}-D^{2}\delta\sigma\end{equation}
Taking the variation to be the {}``time''-derivative, i.e. $\delta\sigma_{ab}=\dot{\sigma}_{ab}=2K_{ab}$
the above becomes\begin{eqnarray}
\delta\tilde{R} & = & -2K^{ab}\tilde{R}_{ab}+2D_{a}D_{b}K^{ab}-2D^{2}K\nonumber \\
 & = & -K\tilde{R}+2D_{a}(D_{b}K^{ba}-D^{a}K)\end{eqnarray}
which is zero if the constraints are fulfilled initially. 

Let us now turn to the variation of the second constraint \begin{eqnarray}
\delta(D_{a}K^{a}\,_{b}-D_{b}K) & = & \delta D_{a}K^{a}\,_{b}+D_{a}\delta K^{a}\,_{b}-D_{b}\delta K\\
 & = & \delta C^{a}\,_{ma}K^{m}\,_{b}-\delta C^{m}\,_{ba}K^{a}\,_{m}+D_{a}\delta K^{a}\,_{b}+D_{b}(K^{mn}K_{mn})\nonumber \end{eqnarray}
in order to evaluate the variation of $K^{a}\,_{b}$ we have to make
use of the constraint to express it completely in terms of $K$, whose
time-variation is given. Since the two-dimensional slices are flat
we may Fourier-transform the constraint, which turns the differential
equation into an algebraic one. Its solution is given by \begin{equation}
K^{a}\,_{b}=D^{a}D_{b}\frac{1}{D^{2}}K\end{equation}
where $1/D^{2}$ denotes the inverse of the Laplacian $D^{2}$. Using
this expression let us first calculate the variation of $K^{a}\,_{b}$\begin{eqnarray}
\delta K^{a}\,_{b} & = & -\delta\sigma^{ac}D_{c}D_{b}\frac{1}{D^{2}}K+\sigma^{ac}\delta D_{c}D_{b}\frac{1}{D^{2}}K\nonumber \\
 &  & -D^{a}D_{b}\frac{1}{D^{2}}\delta D^{2}\frac{1}{D^{2}}K+D^{a}D_{b}\frac{1}{D^{2}}\delta K\nonumber \\
 & = & -2K^{ac}K_{cb}-\delta C^{m}\,_{b}\,^{a}D_{m}\frac{1}{D^{2}}K+D^{a}D_{b}(\frac{1}{D^{2}}(\delta\sigma^{cd}D_{c}D_{d}\frac{1}{D^{2}}K))\nonumber \\
 &  & +D^{a}D_{b}\frac{1}{D^{2}}(\sigma^{cd}\delta C^{m}\,_{dc}D_{m}\frac{1}{D^{2}}K)-D^{a}D_{b}\frac{1}{D^{2}}(K^{cd}K_{cd})\end{eqnarray}
Taking into account that the difference tensor $\delta C^{a}\,_{bc}$
which determines the variation of the derivative operator $D_{a}$
is completely determined by the variation of the metric $\delta\sigma_{ab}$\begin{eqnarray}
\delta C^{a}\,_{bc} & = & \frac{1}{2}(D_{b}\delta\sigma^{a}\,_{c}+D_{c}\delta\sigma^{a}\,_{b}-D^{a}\delta\sigma_{bc})\nonumber \\
 & = & (D_{b}K^{a}\,_{c}+D_{c}K^{a}\,_{b}-D^{a}K_{bc})=D_{b}K^{a}\,_{c}\end{eqnarray}
(where the last equality took the explicit form of $K^{a}\,_{b}$
in terms of $K$ into account) the above becomes\begin{eqnarray}
 & = & -2K^{ac}K_{cb}-D_{b}K^{ma}D_{m}\frac{1}{D^{2}}K+2D^{a}D_{b}\frac{1}{D^{2}}(K^{cd}D_{c}D_{d}\frac{1}{D^{2}}K)\nonumber \\
 &  & +D^{a}D_{b}\frac{1}{D^{2}}(D_{c}K^{mc}D_{m}\frac{1}{D^{2}}K)-D^{a}D_{b}\frac{1}{D^{2}}(K_{cd}K^{cd})\nonumber \\
 & = & -2K^{ac}K_{cb}-D^{m}K^{a}\,_{b}D_{m}\frac{1}{D^{2}}K+D^{a}D_{b}\frac{1}{D^{2}}(K^{cd}K_{cd})\nonumber \\
 &  & +D^{a}D_{b}\frac{1}{D^{2}}(D^{m}KD_{m}\frac{1}{D^{2}}K)\end{eqnarray}
Taking this result into account the variation of $D_{a}K^{a}\,_{b}-D_{b}K$
becomes\begin{eqnarray}
\delta(D_{a}K^{a}\,_{b}-D_{b}K) & = & \delta C^{a}\,_{ma}K^{m}\,_{b}-\delta C^{m}\,_{ba}K^{a}\,_{m}+D_{a}\delta K^{a}\,_{b}-D_{b}\delta K\nonumber \\
 & = & D_{m}KK^{m}\,_{b}-D_{b}K^{m}\,_{a}K^{a}\,_{m}-2D_{a}(K^{ac}K_{cb})\nonumber \\
 &  & -D_{a}(D^{m}K^{a}\,_{b}D_{m}\frac{1}{D^{2}}K)+2D_{b}(K^{cd}K_{cd})\nonumber \\
 &  & +D_{b}(D_{m}K\frac{1}{D^{2}}D^{m}K)\nonumber \\
 & = & 2D_{m}KK^{m}\,_{b}-2D_{b}K^{m}\,_{a}K^{a}\,_{m}-2D_{a}(K^{ac}K_{cd})\nonumber \\
 &  & +2D_{b}(K^{cd}K_{cd})\\
 & = & 0\nonumber \end{eqnarray}

\section{Hamiltonian dynamics}

Since the Einstein-Hilbert action vanishes identically for pp-waves,
which follows from $R_{ab}\propto p_{a}p_{b}$, the question about
a Hamiltonian description does not seem to be a very sensible one.
Nevertheless since the dynamical equations are non-trivial they may
be taken as a starting point for the construction of symplectic structure
as well as a Hamiltonian. In order to exhibit this point of view more
explicitly let us consider electrodynamics first, i.e. try to construct
a Hamiltonian description by starting from the Maxwell equations rather
than the electromagnetic action. 

The source-free Maxwell system\begin{eqnarray}
\epsilon^{abc}D_{b}B_{c}-\dot{E}^{a}=0 &  & D_{a}B^{a}=0\nonumber \\
\epsilon^{abc}D_{b}E_{c}+\dot{B}^{a}=0 &  & D_{a}E^{a}=0\end{eqnarray}
neatly splits into evolution and constraint equations. Introducing
the vector potential $A_{a}$, which we will take as configuration
variable \begin{equation}
B^{a}=\epsilon^{abc}D_{b}A_{c}\end{equation}
solves the first constraint, at the price of being not unique. I.e.
\begin{equation}
A_{a}\longrightarrow A_{a}+D_{a}\Lambda\end{equation}
describes the same physical situation. In order to find the corresponding
momentum we will take a little {}``quantum''--detour. 

Let us assume that the (physical) wave-function $\Psi[A_{a}]$ is
invariant%
\footnote{This is actually a rather strong requirement, but it suffices for
our purpose to identify the canonical momentum%
} under gauge transformations, i.e.\begin{equation}
\Psi[A_{a}+D_{a}\Lambda]=\Psi[A_{a}]\end{equation}
which, by the arbitrariness of $\Lambda$, is equivalent to\begin{equation}
D_{a}\frac{\delta\Psi}{\delta A_{a}}=0.\end{equation}
Identifying the derivative with respect to the configuration variable
(up to a factor $1/i$) with the momentum(operator) suggests to identify
the latter with $E^{a}$. Since we now have derived {}``position''
and {}``momentum'' variables we have constructed the symplectic
form. 

All that is left is to show that the evolution equations are Hamiltonian
with respect to this symplectic form. From\begin{equation}
\dot{E}^{a}=-\frac{\delta H}{\delta A_{a}}=\epsilon^{abc}D_{b}(\epsilon_{cmn}D^{m}A^{n})\end{equation}
we find\begin{eqnarray}
\delta_{A}H & = & -\int\delta A_{a}\epsilon^{abc}D_{b}(\epsilon_{cmn}D^{m}A^{n})\omega_{\delta}\nonumber \\
 & = & -\int\epsilon^{cba}D_{b}\delta A_{a}\epsilon_{cmn}D^{m}A^{n}\omega_{\delta}\nonumber \\
 & = & -\delta\frac{1}{2}\int B_{a}B^{a}\omega_{\delta}\end{eqnarray}
($\omega_{\delta}$ denotes the volume form of $\mathbb{R}^{3}$)
Whereas \begin{eqnarray}
\dot{B}^{a} & = & \epsilon^{abc}D_{b}\dot{A}_{c}=-\epsilon^{abc}D_{b}E_{c}\nonumber \\
0 & = & \epsilon^{abc}D_{b}(\dot{A}_{c}+E_{c})\end{eqnarray}
entails\begin{equation}
\dot{A}_{a}=\frac{\delta H}{\delta E^{a}}=-E_{a}+D_{a}\Lambda\end{equation}
where the last term arises from the kernel of $\epsilon^{abc}D_{b}$.
Upon integration this yields\begin{equation}
\delta_{E}H=-\delta\int(\frac{1}{2}E^{a}E_{a}+D_{a}E^{a}\Lambda)\omega_{\delta}\end{equation}
Putting everything together we find for the Hamiltonian of the Maxwell
system \begin{equation}
H=-\frac{1}{2}\int(E_{a}E^{a}+B_{a}B^{a}+D_{a}E^{a}\Lambda)\end{equation}
which is the {}``correct'' result, i.e. the one obtained from starting
with the electromagnetic action. 

Let us now apply this procedure to the pp-wave system\begin{eqnarray}
\dot{K}+K_{ab}K^{ab}=0 &  & D_{a}K^{a}\,_{b}-D_{b}K=0\nonumber \\
\dot{\sigma}_{ab}=2K_{ab} &  & \tilde{R}=0\end{eqnarray}
The situation is very similar to the electromagnetic case. Again the
system splits into dynamical and constraint equations. Therefore in
the first step we will proceed by trying to identify the symplectic
form. Let us begin by taking the 2-metric $\sigma_{ab}$ as configuration
variable (which is a step motivated from standard 3+1 ADM decomposition).
In order to find the corresponding momentum we will require that the
wave-function should be invariant under (infinitesimal) two-dimensional
diffeomorphisms $\xi^{a}$, i.e.\begin{equation}
\Psi[\sigma_{ab}+D_{a}\xi_{b}+D_{b}\xi_{a}]=\Psi[\sigma_{ab}]\end{equation}
This entails, due to the arbitrariness of $\xi^{a}$\begin{equation}
D_{a}\frac{\delta\Psi}{\delta\sigma_{ab}}=0.\end{equation}
Once again, since the derivative with respect to the configuration
variable (up to a factor $1/i$) represents the momentum(operator)
$\tilde{\pi}^{ab}$ this suggests to identify the latter with\begin{equation}
\tilde{\pi}^{ab}=\omega_{\sigma}(K^{ab}-\sigma^{ab}K)\end{equation}
 if we take the first constraint into account. (Note that momentum
has to be tensor-valued 2-form, which can easily be seen from it being
the derivative of the scalar $\Psi$ with respect to the tensor $\sigma_{ab}$.
In the following the two-form indices will be suppressed in favor
of a tilde). Having identified position and momentum variables, which
is equivalent to the identification of the symplectic structure it
remains to show that the evolution relative to this symplectic structure
is Hamiltonian. Taking into account that\begin{equation}
\tilde{\pi}=-\omega_{\sigma}K\qquad\tilde{\pi}:=\sigma_{ab}\tilde{\pi}^{ab}\end{equation}
the dynamical equations become \begin{eqnarray}
\dot{\tilde{\pi}} & = & \omega_{\sigma}^{-1}(\tilde{\pi}^{ab}\tilde{\pi}_{ab}-\tilde{\pi}^{2})\nonumber \\
\dot{\sigma}_{ab} & = & 2\omega_{\sigma}^{-1}(\tilde{\pi}_{ab}-\sigma_{ab}\tilde{\pi})\end{eqnarray}
in terms of the canonical variables. (Here the expression $\omega_{\sigma}^{-1}$
denotes the inverse volume form of the $2$-slice, i.e. locally $\omega_{\sigma}^{-1}=1/\sqrt{\sigma}\partial_{1}\wedge\partial_{2}$)
Integration of the second equation of motion gives \begin{equation}
\dot{\sigma}_{ab}=\frac{\delta H}{\delta\tilde{\pi}^{ab}}\qquad\delta_{\pi}H=\int2\omega_{\sigma}^{-1}\delta\tilde{\pi}^{ab}(\tilde{\pi}_{ab}-\sigma_{ab}\tilde{\pi})=\delta_{\pi}\int\omega_{\sigma}^{-1}(\tilde{\pi}^{ab}\tilde{\pi}_{ab}-\tilde{\pi}^{2}).\end{equation}
Let us now derive the variation of $\pi$. Taking into account that
$\tilde{\pi}=\sigma_{ab}\tilde{\pi}^{ab}$ we have\begin{eqnarray}
\dot{\tilde{\pi}} & = & \dot{\sigma}_{ab}\tilde{\pi}^{ab}+\sigma_{ab}\dot{\tilde{\pi}}^{ab}=2\omega_{\sigma}^{-1}(\tilde{\pi}^{ab}\tilde{\pi}_{ab}-\tilde{\pi}^{2})-\sigma_{ab}\frac{\delta H}{\delta\sigma_{ab}}\nonumber \\
 & = & 2\omega_{\sigma}^{-1}(\tilde{\pi}^{ab}\tilde{\pi}_{ab}-\tilde{\pi}^{2})-\sigma_{ab}(2\omega_{\sigma}^{-1}(\tilde{\pi}^{ac}\tilde{\pi}_{c{}}\,^{b}-\tilde{\pi}^{ab}\tilde{\pi})-\frac{1}{2}\omega_{\sigma}^{-1}\sigma^{ab}(\tilde{\pi}^{cd}\tilde{\pi}_{cd}-\tilde{\pi}^{2}))\nonumber \\
 & = & \omega_{\sigma}^{-1}(\tilde{\pi}^{ab}\tilde{\pi}_{ab}-\tilde{\pi}^{2})\end{eqnarray}
where the expression for $H$ has been taken from the previous. Since
the result coincides with the first equation of motion we may take
\begin{equation}
H[\sigma,\tilde{\pi}]=\int\omega_{\sigma}^{-1}(\tilde{\pi}^{ab}\tilde{\pi}^{cd}\sigma_{ac}\sigma_{bd}-(\tilde{\pi}^{cd}\sigma_{cd})^{2})\end{equation}
to be the Hamiltonian of the our system.

\section*{Conclusion}

We have shown that it is possible to formulate the dynamics of the
pp-wave system similar to the Gaussian evolution of the standard timelike
situation. At first sight the vanishing action, i.e. its topological
nature, seems to hamper a Hamiltonian formulation. Nevertheless upon
comparison with the electromagnetic system we succeed in identifying
both symplectic structure as well as the dynamical Hamilton function.
We believe that this opens the road to the quantization of the model
in terms of a midi-superspace formulation. Work in this direction
is currently in progress.

\end{document}